\documentclass[pdflatex,
sn-nature,
]{sn-jnl}

\usepackage[style=nature, defernumbers=true, backend=biber]{biblatex}

\DeclareBibliographyCategory{mainpart}
\usepackage{etoolbox}
\newtoggle{inmainpart}  
\AtEveryCitekey{%
  \iftoggle{inmainpart}{%
    \addtocategory{mainpart}{\thefield{entrykey}}
  }{}%
}
\newenvironment{mainpart}{%
  \begingroup
  \toggletrue{inmainpart}
}{%
  \endgroup  
}
\let\oldcite\cite
\renewcommand{\cite}[1]{\textsuperscript{\oldcite{#1}}}
\let\cite\supercite

\usepackage{graphicx}
\usepackage{xcolor}
\usepackage{amsmath,amssymb,amsfonts}

\usepackage{float} 
\newfloat{extendedFig}{htbp}{loflow}
\floatname{extendedFig}{Extended Data Fig.} 

\usepackage{siunitx} 

\begin{document}

\title{Attosecond Control and Measurement of Chiral Photoionisation Dynamics}

\author*[1,2]{\fnm{Meng} \sur{Han}}\email{meng.han@phys.chem.ethz.ch}
\equalcont{These authors contributed equally to this work.}

\author[1]{\fnm{Jia-Bao} \sur{Ji}}
\equalcont{These authors contributed equally to this work.}

\author[3]{\fnm{Alexander} \sur{Blech}}

\author[4]{\fnm{R. Esteban} \sur{Goetz}}

\author[2]{\fnm{Corbin} \sur{Allison}}

\author[2]{\fnm{Loren} \sur{Greenman}}

\author[3]{\fnm{Christiane P.} \sur{Koch}}

\author*[1]{\fnm{Hans Jakob} \sur{W\"orner}}\email{hwoerner@ethz.ch}

\affil[1]{Laboratorium für Physikalische Chemie, ETH Zürich, Zürich, 8093, Switzerland}

\affil[2]{J. R. Macdonald Laboratory, Department of Physics, Kansas State University, Manhattan, KS 66506, USA}

\affil[3]{Institut f{\"u}r Theoretische Physik, Freie Universit{\"a}t Berlin, Berlin, 14195, Germany}

\affil[4]{Department of Physics, University of Connecticut, Storrs, CT 06269, USA}


\maketitle

\clearpage

\begin{mainpart}

\textbf{
Many chirality-sensitive light-matter interactions are governed by chiral electron dynamics. Therefore, the development of advanced technologies harnessing chiral phenomena would critically benefit from measuring and controlling chiral electron dynamics on their natural attosecond time scales. Such endeavors have so far been hampered by the lack of characterized circularly polarized attosecond pulses, an obstacle that has recently been overcome \cite{Han2023metrology,Han2023attosecond}. In this article, we introduce chiroptical spectroscopy with attosecond pulses and demonstrate attosecond coherent control over photoelectron circular dichroism (PECD)\cite{goetz2019quantum,goetz2024rabbitt}, as well as the measurement of chiral asymmetries in the forward-backward and angle-resolved photoionisation delays of chiral molecules. 
We show that co-rotating attosecond and near-infrared pulses can nearly double the PECD and even change its sign compared to single-photon ionisation. We demonstrate that chiral photoionisation delays depend on both polar and azimuthal angles of photoemission in the light-propagation frame, requiring three-dimensional momentum resolution. We measure forward-backward chiral-sensitive delays of up to 120~as and polar-angle-resolved photoionisation delays up to 240~as, which include an asymmmetry of $\sim$60~as originating from chirality in the continuum–continuum transitions. Attosecond chiroptical spectroscopy opens the door to quantitatively understanding and controlling the dynamics of chiral molecules on the electronic time scale.
}

Chirality is traditionally viewed as a structural property of matter. The spatial arrangement of atoms in molecules and materials indeed defines their handedness. In a simplified view, structural chirality is often used to explain chiral recognition. 
However, recent research indicates that structural chirality is insufficient to fully understand chiral phenomena. In particular, growing evidence points to a functional role of chiral electron dynamics in enabling and mediating chiral interactions at the electronic level \cite{bloom24a}. Such dynamics are also thought to play a role in advanced applications including spintronics \cite{suda19a}, unidirectional molecular machines \cite{zhu21a}, and chiral-sensitive biosensing \cite{ziv21a}. Furthermore, a wide range of chiroptical techniques, such as electronic circular dichroism \cite{rodgers12a} (CD), photoelectron circular dichroism \cite{ritchie1976theory, bowering2001} (PECD), and chiral high-harmonic generation \cite{Cireasa2015,Baykusheva2018}, are now understood to probe the underlying chiral dynamics of electrons rather than merely structural asymmetries. These developments motivate a broader perspective on chirality that encompasses not only static structure but also the dynamic behavior of electrons in chiral systems.

In spite of the fundamental importance of chiral electron dynamics, spectroscopy based on attosecond pulses has so far been lacking the crucial capability of chiral discrimination. The key obstacle for expanding attosecond science into the realms of chiral molecules and materials has 
been the lack of circularly polarized attosecond light pulses. As a consequence, all pioneering experiments in this field have relied on femtosecond pulses, e.g. Refs. \cite{Lux2012,Lehmann2013,Lux2015,Cireasa2015,Beaulieu2017,beaulieu18a,Baykusheva2018,Baykusheva2019,rozen19a,Svoboda2022}. 
Notably, strong-field ionisation with intense two-color femtosecond laser pulses has been used to measure phase shifts \cite{Beaulieu2017} and control the asymmetry of photoelectron angular distributions (PADs) in the strong-field ionisation of chiral molecules \cite{demekhin18a,rozen19a}.
Recently, some of the present authors have developed attosecond metrology in circular polarization by introducing a plug-in apparatus for the generation and demonstrating a general methodology for the complete characterization of circularly polarized attosecond pulses \cite{Han2023metrology}. This new capability has recently been applied to study the photoionisation dynamics of atoms \cite{Han2023attosecond,han2024separation}. 

Here, we introduce attosecond chiroptical spectroscopy, by reporting the first application of circularly polarized attosecond pulses to chiral molecules. This development, combined with momentum-vector-resolved electron-ion-coincidence spectroscopy, allows us to measure and control chiral electron dynamics on their natural attosecond time scale. Specifically, our study demonstrates attosecond coherent control over PECD based on the constructive or destructive interference between pairs of well-defined photoionisation pathways. It also reveals the characteristic dependence of chiral photoionisation delays on both the polar and the azimuthal angles of photoemission in the light-propagation frame.

The experimental setup is schematically illustrated in Fig. \ref{fig:figure1}a, with details given in the Methods section. The circularly polarized extreme-ultraviolet (XUV) attosecond pulse train (APT) of a single, common helicity was produced by the non-collinear HHG scheme using a compact plug-in apparatus \cite{Han2023metrology}. A weak linearly or circularly polarized (co-rotating) near-infrared (IR, 800 nm) laser pulse was spatio-temporally overlapped with the XUV APT with attosecond temporal stability to induce XUV/IR two-photon transitions using the reconstruction of attosecond beating by interference of two-photon transitions (RABBIT) approach. In this study, we define the light propagation direction to be the $z$ axis and the light polarization to be in the $xy$ plane, so that the CD manifests itself in the angular distribution with respect to the angle $\theta = {\rm{arccos}}(p_z/p_{\rm total})$, where $p_{\rm total}$ and $p_z$ are the total momentum and its $z$ component, respectively. The azimuthal angle in the polarization plane $\phi = {\rm{atan}}(p_x/p_y)$ was integrated over from $-15^\circ$ to $+15^\circ$. 
Enantiomerically pure samples of methyloxirane (MeOx, also known as propylene oxide, C$_3$H$_6$O, $\geq$99.5\% ee) were delivered into a cold-target recoil-ion-momentum spectroscopy (COLTRIMS) setup \cite{Dorner2000, Ullrich2003} by supersonic expansion through the nozzle with an aperture of 30 $\mu$m. The three-dimensional (3D) momenta of photoelectrons and ionic fragments were measured in coincidence in the ultrahigh vacuum chamber ($\sim 10^{-10}$ mbar).

\begin{figure}[t]
\centering
\includegraphics[width=\linewidth]{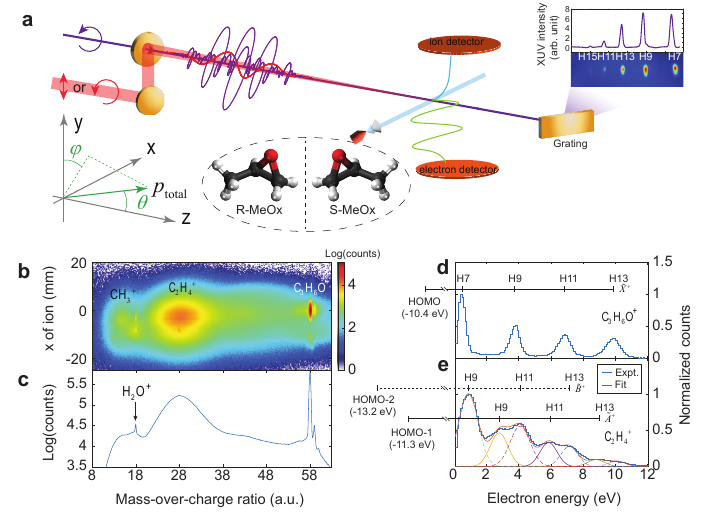}
\caption{\textbf{Attosecond chiroptical coincidence spectroscopy}. \textbf{a}, Experimental setup. The ionic fragments and emitted photoelectrons are measured in coincidence using COLTRIMS. \textbf{b}, Measured two-dimensional ionic spectrum as a function of the mass-over-charge ratio and the position on the ion detector along the molecular beam direction in the XUV field only. \textbf{c}, The projection of the distribution in \textbf{b} to the mass-over-charge axis. Note that the counts are in the logarithmic scale. The two small peaks above $m/q= 58$ correspond to $^{13}$C isotopomers of the parent ion. \textbf{d, e}, Measured photoelectron spectra in coincidence with the intact parent ion (\textbf{d}) and the $\rm{{C_2H_4}^+}$ fragment ion (\textbf{e}). In \textbf{d}, a series of discrete equidistant peaks validates that the photoionisation to the $\widetilde{X}^+$ ground state of the cation leaves the molecule intact. In \textbf{e}, the experimental data is fitted by two sets of Gaussian peaks corresponding to the $\widetilde{A}^+$ and $\widetilde{B}^+$ ionic states, respectively. The ionisation energies of the relevant states are adopted from the reference \cite{garcia2014photoelectron}. }
\label{fig:figure1}
\end{figure}

Figure \ref{fig:figure1}b shows the ionic fragment distribution as a function of mass-over-charge ratio ($m/q$) and detector position, while Fig. \ref{fig:figure1}c presents the projection onto the $m/q$ axis, both recorded using attosecond XUV pulses only. After photoionisation by the attosecond pulses comprising harmonic 7 (H7) to harmonic 13 (H13), we observe the undissociated parent ion ($\rm{C_{3}H_{6}O^{+}}$) and two broader distributions assigned to $\rm{{CH_{3}}^+}$ and $\rm{{C_{2}H_{4}}^{+}}$ fragments, in agreement with previous synchrotron studies \cite{garcia2014photoelectron}. The photoelectron spectrum identifies the electronic state of the ion that was accessed in the ionisation step, whereas the ionic species reveal whether and how a given electronic state of the ion dissociates. Figures \ref{fig:figure1}d and \ref{fig:figure1}e show the photoelectron spectra measured in coincidence with the parent ion and $\rm{{C_2H_4}^+}$, respectively. The former reflects photoionisation from the highest occupied molecular orbital (HOMO) by H7-H13, which leaves the cation intact. The photoelectron spectrum measured in coincidence with $\rm{{C_2H_4}^+}$ shows a more complex distribution, which contains contributions from the $\widetilde{A}^+$ and $\widetilde{B}^+$ ionic states. In this study, we focus on the parent-ion channel.

\begin{figure}[t]
\centering
\includegraphics[width=\linewidth]{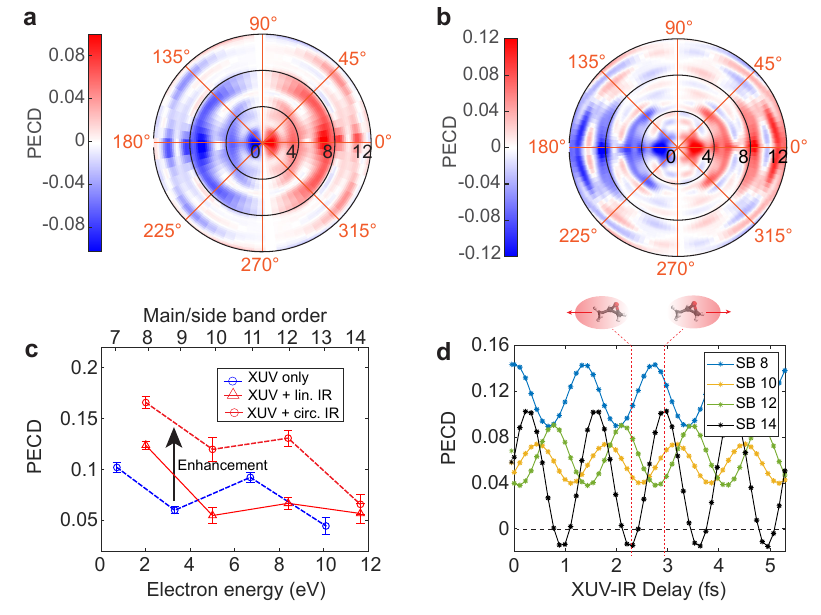}
\caption{\textbf{Two-color enhancement and attosecond coherent control over PECD detected in coincidence with the parent ion}. \textbf{a, b} Measured angle- and energy-resolved PECD distributions in the XUV-only case and the XUV+IR case with linearly polarized IR field, respectively. The symbols on top of the panels label the light polarization states. \textbf{c}, Comparison of PECD values between the XUV-only case and the (delay-averaged) XUV+IR two-color case, including both linear and co-rotating circular polarizations of the IR field. \textbf{d}, Measured PECD values at SBs 8, 10, 12 and 14, as a function of the XUV-IR delay for a linearly polarized IR field. The photoelectron signals have been integrated over a window of 0.5 eV width centered at the peaks of the main bands and SBs. The insets above panel d illustrate the PECD reversal observed in SB14.}
\label{fig:figure2}
\end{figure}

Figure \ref{fig:figure2} shows the measured PECD and its coherent control by changing the XUV-IR delay with attosecond precision for the parent-ion channel. The energy($E_k$)- and angle($\theta$)-resolved PECD distribution is defined by 2[$I_S$($E_k$,$\theta$) - $I_R$($E_k$,$\theta$)]/[$I_S$($E_k$,$\theta$) + $I_R$($E_k$,$\theta$)], where $I_{S/R}$($E_k$,$\theta$) is the PAD for the $S/R$ enantiomer. The chiral nature of PECD has also been verified by switching the XUV helicity. Figures \ref{fig:figure2}a and \ref{fig:figure2}b display the measured PECD distributions obtained from the XUV field only and the delay-averaged result in the XUV+IR two-color fields, respectively. For the XUV-only PECD distribution, the four dipole-shaped concentric rings correspond to the four main peaks shown in Fig. \ref{fig:figure1}d. After introducing an IR field (Fig. \ref{fig:figure2}b), sidebands (SBs) appear in between the main peaks and inherit the PECD-sign of the neighboring main peaks but with some angular modulations, a consequence of the fact that additional partial waves are involved in the two-photon-ionisation process (see Theoretical Methods section for the details). The asymmetry parameters are obtained by fitting spherical-harmonic functions to the angle-resolved PECD with results shown in Extended Data Fig. 1. Figure \ref{fig:figure2}c compares the extracted PECD values in the XUV-only and XUV+IR cases. In the XUV-only case the highest PECD amounts to $\sim$12$\%$. In the presence of a linearly polarized IR field, the PECDs at the SB positions are enhanced by up to three percent compared to the average PECD of the neighboring main bands. In the presence of a circularly polarized IR field, the SB PECDs are increased much more, which is in line with predictions from Ref.~\cite{goetz2024rabbitt} and simulations performed for MeOx. Typically, the PECD values are doubled compared to the average of their neighboring main bands, as indicated by the black arrow in Fig. \ref{fig:figure2}c. In the XUV+IR field, the PECD can additionally be actively controlled. Figure \ref{fig:figure2}d shows the delay-resolved PECD in the presence of a linearly polarized IR field. The PECDs at all SB positions are modulated with a period of 1.33 fs, with a maximal modulation depth at SB 14, where the PECD even changes sign. The phase shifts between the PECD oscillations of different SBs are dominated by the attochirp. The delay-dependent PECD in the case of a circularly polarized IR field is shown in Extended Data Fig. \ref{fig:figure2}.

\begin{figure}[t]
\centering
\includegraphics[width=\linewidth]{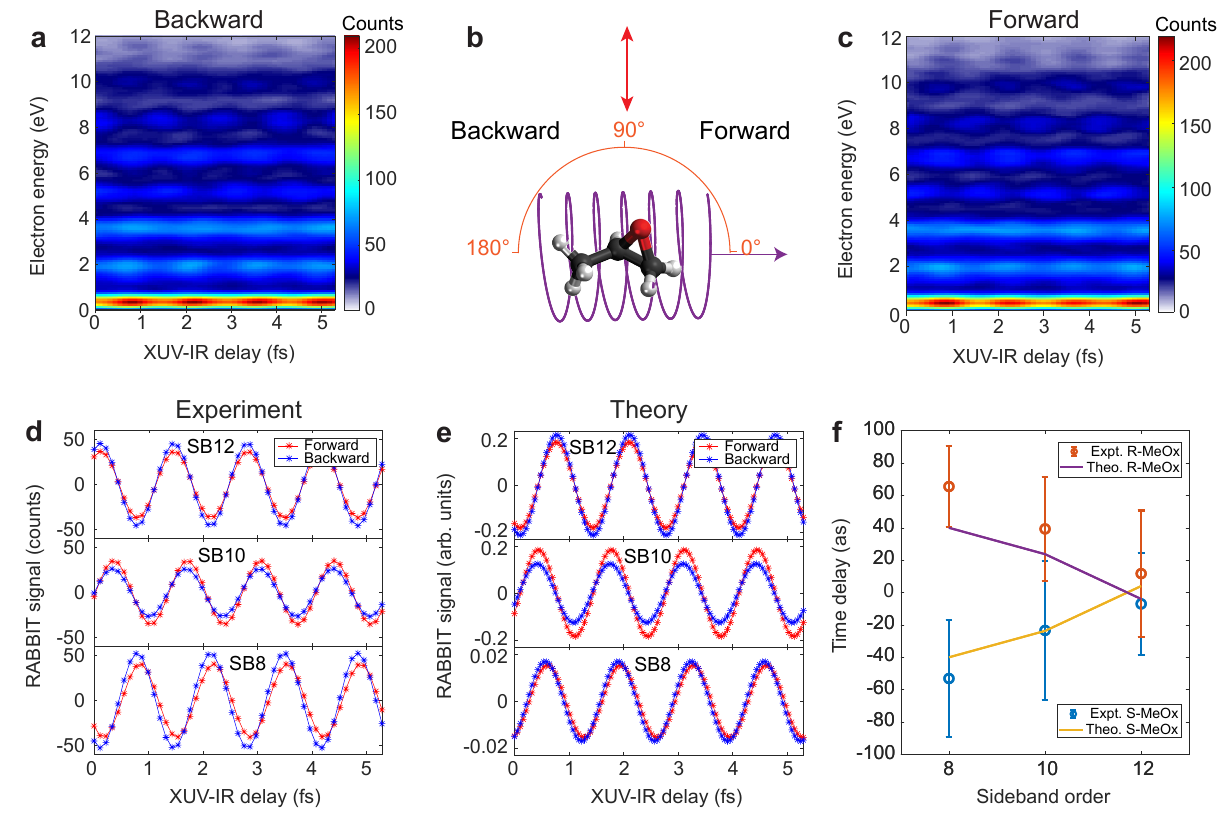}
\caption{\textbf{Chiral asymmetries in photoionisation time delays}. Data were acquired with a linearly polarized IR along the y direction, and the photoemission $\phi$ angle was integrated over from $-15^\circ$ to $+15^\circ$. (\textbf{a} and \textbf{c}) RABBIT traces of backward electrons ($\theta$ is integrated over from $90^\circ$ to $180^\circ$) and forward electrons ($\theta$ is integrated over from 0 to $90^\circ$), respectively, as shown in panel (\textbf{b}). (\textbf{d} and \textbf{e}) Measured and simulated energy-integrated RABBIT signals for SBs 8, 10 and 12, where the constant background signals have been removed using Fourier transformation. The energy integration width is 0.5 eV centered at the peak positions. (\textbf{f}) Experimental and theoretical forward-backward photoionisation time delay as a function of photoelectron energy for both enantiomers of MeOx. Note that the experimental results in panels (a,c,d,e) are corresponding to the $R$-MeOx, and the data for the $S$-MeOx is shown in Extended Data Fig. 3.}
\label{fig:figure3}
\end{figure}

We now discuss the measurement of chiral asymmetries in molecular photoionisation delays, starting with the case of a linearly polarized IR field. As in the case of PECD, this approach isolates the chiral signatures of the one-photon XUV transition and eliminates possible chiral contributions from continuum-continuum transitions driven by the IR field \cite{goetz2024rabbitt}. Figures \ref{fig:figure3}a and \ref{fig:figure3}c display the energy-resolved RABBIT traces obtained by integrating over the photoelectron emission angle $\theta$ from 0° to 90° (forward-emitted photoelectrons) or from 90° to 180° (backward-emitted photoelectrons). The phases of the photoelectron-yield oscillations at the SB positions have contributions from both the attochirp and the molecular photoionisation delays. By evaluating the phase difference between forward- and backward-emitted photoelectrons, the influence of the attochirp is cancelled, isolating the chiral asymmetry of the molecular photoionisation delays. Figure \ref{fig:figure3}d compares the yield oscillations of forward and backward photoelectrons for different SBs, after removal of the slowly varying background by Fourier transformation (see Methods for details). The forward photoelectrons are temporally behind the backward photoelectrons, i.e. the SB maxima occur at larger XUV-IR delays, an effect that is most prominent for SB8. The same measurements performed on the other enantiomer ($S$-MeOx, see Extended Data Fig. 3), revealed opposite delays, demonstrating their chiral nature. Figure \ref{fig:figure3}f shows the measured photoionisation time delays for both enantiomers. We find that the forward-backward time delays, expressing the temporal separation of the forward- and backward-emitted photoelectron wave packets, amounts to $\sim$60 as at SB8 and decreases with electron kinetic energy, which reflects the decreasing sensitivity of the photoelectron wavepacket to the chirality of the molecular potential, as in the case of PECD amplitudes.
The calculated time delays are somewhat smaller than the measured ones, but lie within the error bars of the measurement. The chiral nature of these forward-backward time delays is confirmed by the opposite signs of the time delays measured for the two enantiomers in both experiment and theory.

The molecular photoionisation dynamics are simulated by adapting the theoretical framework of Ref. \cite{goetz2024rabbitt} which suggested to leverage RABBIT to realize the coherent control of chiral signatures in the PAD.
The simulations are performed in the frozen-core static-exchange and electric-dipole approximations, following Refs.~\cite{goetz2019quantum,goetz2019perfect,goetz2024rabbitt} with the modifications described in the Methods section to remedy the effect of artificial resonances caused by the discretization of the photoelectron continuum and to account for the experimental conditions.

The original theory proposal discussed photoelectron interferometry for the model chiral molecule CHBrClF \cite{goetz2024rabbitt}. Compared to CHBrClF, MeOx shows a more isotropic nature, as can be seen from the smaller number of angular basis functions needed to represent the molecular wavefunctions for MeOx compared to CHBrClF (see Methods). This suggests that the photoelectrons of CHBrClF experience higher anisotropy in the molecular potential.
To what extent does this influence the robustness of the control scheme?
Simulations for CHBrClF in Ref.~\cite{goetz2024rabbitt}, showed that the maximum PECD signal could be enhanced by a factor of five by optimizing the XUV-IR delay, while in the simulations for MeOx optimizing the delay can increase the PECD by about a factor of two, which is a similar enhancement as the delay-averaged experimental results presented in Fig.~\ref{fig:figure3}.
This may indicate that the presented control scheme could benefit from a less isotropic nature of the molecule.
A similar argument can be made when analysing the dependence of the chiral signatures on the photoelectron energy.
The PECD as well as the forward-backward time delays decrease with the SB order as shown in Fig.~\ref{fig:figure2}c and Fig.~\ref{fig:figure3}f.
Nevertheless, introducing the circularly polarized IR pulse increases PECD by about 50 \% for the highest SB (SB14) compared to the PECD signal of the closest harmonic (H13) in the XUV-only case.

\begin{figure}[t]
\centering
\includegraphics[width=\linewidth]{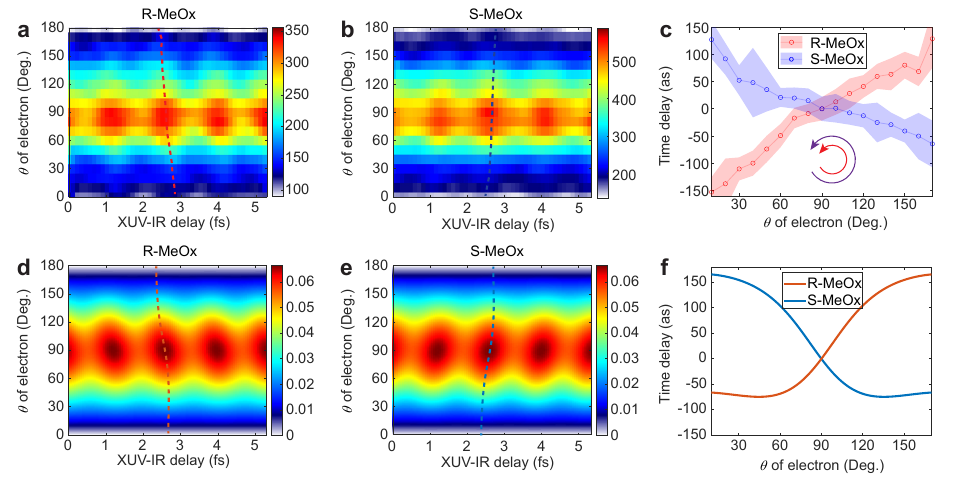}
\caption{\textbf{Angle-resolved photoionisation time delays in co-rotating XUV+IR fields}. (\textbf{a} and \textbf{b}) $\theta$-resolved RABBIT traces of SB 8 for the $R$ and $S$ enantiomers of MeOx. (\textbf{c}) The extracted photoionisation time delay from a and b, where the delay values at $\theta = 90^\circ$ were chosen as reference. Note the negative value of the ionisation time delay represents the electron is delayed with respect to our reference of $\theta = 90 ^\circ$. These two RABBIT-phase curves are also plotted in a and b as dashed lines, and the equivalent lines are shown in d and e. The uncertainty of the extracted time delay (error bar) is estimated by the background-over-signal approach \cite{Ji2021}. (\textbf{d-f}) The corresponding results from theoretical calculations described in the text and Methods Section.}
\label{fig:figure4}
\end{figure}

In addition to the chiral forward-backward time delays, the 3D momentum resolution of our experiment allows us to obtain the photoionisation time delays with angular resolution. Momentum-vector resolution is a prerequisite for a quantitative measurement of chiral photoionisation delays because the latter depend on both laboratory-frame angles of photoemission ($\theta$ and $\phi$) in any chiral-sensitive experiment. Extended Data Fig. 4 illustrates this fact by showing that the phase of the SB oscillation linearly depends on $\phi$ in the case of co-rotating XUV and IR fields. This is the RABBIT analogue of the angular streaking principle. Taking this property into account, allows us to perform a quantitative analysis of angle-resolved chiral photoionisation delays.
Figures 4a and 4b display the measured $\theta$-resolved RABBIT traces of SB8 for the two enantiomers of MeOx in the case of a co-rotating circularly polarized IR field, the configuration that yields the largest PECD enhancement in line with theoretical calculations. The RABBIT fringes are not vertical, but display a tilt as a function of $\theta$, in a direction that inverts when exchanging the enantiomers, again demonstrating the chiral nature of the effect. The time delays extracted at each $\theta$ angle (relative to $\theta=90^\circ$) are shown in Fig. \ref{fig:figure4}c. They vary by $\sim$240~as over the probed angular range. Figures \ref{fig:figure4}d and \ref{fig:figure4}e show the calculated angle-resolved RABBIT traces for the two enantiomers, respectively, nicely reproducing the opposite tilts of the RABBIT fringes for the two enantiomers. The angle-resolved photoionisation delays shown in Fig. \ref{fig:figure4}f nicely reproduce the measured variation of the photoionisation delays by $\sim$240~as over the measured angular range. The remaining differences between the theoretical results and the experimental data may originate from non-dipole effects, non-perturbative effects in the light-molecule interaction or electronic correlation effects. The experiment was performed well within the perturbative regime and the results shown in Fig. \ref{fig:figure2}a,b do not show evidence for strong non-dipole effects. We therefore expect electronic correlations to be the primary cause for deviations between experiment and simulations.

The comparison of angle-resolved time delays obtained with circularly or linearly polarized IR pulses additionally provides insights into the contribution of the continuum-continuum transitions to the chiral asymmetries of the time delays. Extended Data Fig. 5 shows such a direct comparison with panels a and b displaying the raw experimental data for the case of a linearly polarized IR field, whereas panels c and d compare the angle-dependent photoionisation time delays for linearly and circularly polarized IR fields, respectively. Whereas the delay variation amounts to $\sim$180 as in the case of a linear IR field, it reaches $\sim$240 as in the case of a circularly polarized IR field. To quantify the chiral asymmetry contributed by the continuum-continuum transitions, panel e shows the symmetrized time delays for a given polarization configuration (obtained as $(\tau(\theta)_{S-\rm{MeOx}} + \tau(180^\circ-\theta)_{R-\rm{MeOx}})/2$) and panel f shows their difference, which quantifies the chiral contribution of the continuum-continuum transitions to the measured photoionisation delays. Although the error ranges overlap with zero, a clear trend can be recognized, corresponding to a contribution of $\sim$60 as that originates from the chirality of the IR-induced continuum-continuum transitions in the chiral potential.

A comparison of the present results, obtained with circularly polarized attosecond pulses in the perturbative regime, with previous results, obtained with femtosecond pulses in the strong-field regime \cite{Beaulieu2017,demekhin18a,rozen19a}, further highlights the achieved advances. The perturbative nature of the present approach results in a transparent control mechanism and good agreement with calculations. In the present work, the naturally occurring PECD effect has been enhanced by a factor of two, reaching up to 16\%, and has been coherently controlled on the attosecond time scale in the perturbative regime involving pathways driven by one XUV and one IR photon. Using femtosecond pulses, Rozen et al. have observed asymmetries of up to 0.5 \% and their coherent control using non-perturbative two-color strong-field ionisation, but a detailed understanding of the mechanism as well as agreement with theory were missing \cite{rozen19a}. In this work, we have measured chirality-sensitive photoionisation delays by advancing the established RABBIT technique to circularly polarized attosecond pulses. The photoionisation delays have been measured with three-dimensional momentum resolution, in coincidence with specific ionic fragments and resolved in terms of the cationic final states. Good agreement with theory and a transparent explanation of the underlying mechasnisms have been provided. In contrast, Beaulieu et al. \cite{Beaulieu2017} have used strong-field ionisation based on non-perturbative two-color femtosecond pulses. They have measured the projection of the PAD on a two-dimensional detector parallel to the propagation direction of the laser pulses. This geometry does not resolve the emission angle in the polarization plane of the laser pulses on which the photoionisation delay depends. Moreover, the photoionisation delays determined in two-color strong-field ionisation depend on the intensity of the laser pulses \cite{zipp2014}, such that they do not represent intrinsic molecular properties.

In conclusion, we have introduced attosecond chiroptical spectroscopy and have applied it to resolve and control the photoionisation dynamics of chiral molecules on their natural, attosecond time scale. Our study demonstrates the powerful capability of the combination of circularly polarized attosecond light pulses with the 3D-momentum-resolved electron-ion coincidence detection, which allows to probe the attosecond electron dynamics induced by photoionisation in an electronic-state-, angle-, energy- and enantiomer-resolved manner. Electron scattering in a chiral molecular potential gives rise to the asymmetric scattering amplitudes with respect to the light propagation direction, underlying the PECD phenomenon, which has high chiral sensitivity and broad applicability. Attosecond chiroptical spectroscopy has now additionally revealed that one-photon-ionisation delays of chiral molecules also display chiral asymmetries, defining a general approach to time resolve the electron scattering dynamics in chiral molecules.
Attosecond chiroptical spectroscopy also has the potential of answering a very important question regarding the origin of the chirality-induced spin selectivity (CISS) effect \cite{naaman19a}, which still lacks a quantitative explanation. Whereas traditional approaches view CISS as a purely electronic (spin-orbit) effect, such models underestimate the CISS effect by typically two orders of magnitude \cite{evers22a}. Very recent models suggest that coupled electronic-nuclear dynamics could play a central role \cite{wu21a}. Attosecond chiroptical spectroscopy may offer a solution to this intriguing puzzle through its ability of temporally separating electronic from structural dynamics \cite{matselyukh22a}.

\end{mainpart}

\printbibliography[category=mainpart, title={References}]

\section*{Acknowledgments}
We thank A. Schneider and M. Seiler for their technical support. \textbf{Funding} M.H. acknowledges the funding from the European Union’s Horizon 2020 research and innovation program under the Marie Skłodowska-Curie grant agreement No 801459 - FP-RESOMUS. This work was supported by ETH Z\"urich and the Swiss National Science Foundation through projects 200021\_172946 and the NCCR-MUST, by the Deutsche Forschungsgemeinschaft 
Project No. 328961117—SFB 1319 ELCH (Extreme light for sensing and driving molecular chirality).
The computing for this project was performed on the Beocat Research Cluster at 
Kansas State University, which is funded in part by NSF grants CNS-1006860, EPS-1006860, and EPS-0919443, ACI-1440548, CHE-1726332, and NIH P20GM113109,
and used resources of the National Energy Research Scientific
Computing Center (NERSC), a U.S. Department of Energy Office of Science User
Facility operated under Contract No. DE-AC02-05CH11231 using NERSC award BES-ERCAP0024357. M. H., C. A. and L. G. were supported by the Chemical Sciences, Geosciences, and Biosciences Division, Office of Basic Energy Sciences, Office of Science, U.S. Department of Energy, under Grant No. DE-SC0022105. C. A. also acknowledges NSF Grant No. 2244539 for additional support during the summer of 2023. A. B. and C. P. K. acknowledge financial support from the Deutsche Forschungsgemeinschaft (CRC 1319).

\section*{Authors contributions} 
H.J.W and M.H. conceived the study. M.H. and J.-B.J. performed the experiments and analysed the data. Simulations were implemented and carried out by A.B., E.G. and C.A. with the supervision of L.G. and C.P.K.. H.J.W. supervised the experimental part of the project. M.H. and H.J.W. wrote the paper with the input of all co-authors.

\section*{Competing interests} 
The authors declare no competing interests.

\section*{Data availability} 
The data generated and analysed in this study is available on the public ETH data repository with the identifier \texttt{10.3929/ethz-b-000741781}.

\section*{Methods}
\subsection*{Experimental methods}
\noindent\textbf{Experimental setup}. Near-infrared laser pulses (2 mJ) were delivered from a regenerative Ti:sapphire laser amplifier at a central wavelength of 800 nm with the repetition rate of 5 kHz. The pulse duration of 25 fs (full width at half maximum in intensity) was measured by a home-built trasient-grating FROG apparatus. This laser beam was split with a 70:30 beam splitter, and the more intense part was sent through our beam-in-beam module and then focused by a silver mirror ($f$ = 45 cm) into a 3 mm long, xenon-filled gas cell to generate a circularly polarized extreme-ultraviolet attosecond pulse train via high-harmonic generation. Our beam-in-beam module consists of a specially designed half waveplate and a common quarter waveplate. The half waveplate is composed of two half disks and the fast-axis angles of the two parts are oriented at $\pm 22.5^\circ$ with respect to the horizontal boundary, respectively. After passing through this half waveplate, the polarization directions of the upper part and the bottom part of the beam are orthogonal to each other. The two parts of the beam are then converted to be circularly polarized but with opposite helicities by transmission through the quarter waveplate \cite{Han2023metrology}. The left-circularly polarized XUV beam in the two produced beams was picked up using a perforated mirror and focused into the main chamber of a COLTRIMS by a nickel-coated toroidal mirror ($f$ = 50 cm). The XUV spectrum was characterized with a home-built XUV spectrometer consisting of an aberration-corrected flat-field grating (Shimadzu 120 lines/mm) and a micro-channel-plate (MCP) detector coupled to a phosphor screen. The beam with $30\%$ energy was used as the dressing field in RABBIT experiments. The dressing IR field was adjusted to linear polarization or left-circular polarization by a zero-order quarter waveplate and its intensity was controlled at a very low level (about $10^{11}~ \textrm{W/cm}^2$) by an iris. The dressing IR beam was focused by a perforated lens ($f$ = 50 cm) and then recombined with the XUV beam by a perforated mirror. In the arm of the dressing field, there were two delay stages, i.e., a high-precision direct-current motor (PI, resolution 100 nm) and a piezoelectric motor (PI, resolution 0.1 nm), adjusting the interpulse delays on femtosecond and attosecond time scales, respectively. A HeNe continuous-wave laser beam was sent through the beam splitter and traced the IR path in the two arms of the interferometer. A fast CCD camera behind the recombination mirror was used to lock the phase delay between the two arms through a PID feedback. When scanning the XUV-IR phase delay in the measurements, the piezo delay stage was actively stabilized at the step size of 110 as with the time jitter of less than 30 as.

\noindent\textbf{Sample delivery and detection}. The enantiopure MeOx samples were purchased from Sigma-Aldrich and contained in a heatable bubbler ($\sim$ 0.5 bar at 30$^\circ$C). The supersonic gas jet was delivered along the $x$ direction through a small nozzle with an opening hole diameter of 30 $\mu m$ and passed through two conical skimmers (Beam Dynamics) located 10 mm and 30 mm downstream with diameters of 0.2 mm and 1 mm, respectively. For the COLTRIMS spectrometer, static electric ($\sim$ 2.21 V/cm) and magnetic ($\sim$ 5.92 G) fields were applied along the $y$ axis to collect the charged fragments (electrons and ions) in coincidence. We did not observe any Coulomb explosion channels in our photoion-photoion coincidence (PIPICO) spectrum. Only the single ionisation channels (one electron is coincident with one ion) were analysed and presented in this work.

\noindent\textbf{Data Analysis}. 
In circularly polarized IR fields, the photoelectron azimuthal angle plays the same role as the XUV-IR delay \cite{Han2023metrology} due to the principle of angular streaking \cite{zhao05a}, as illustrated in Extended Data Fig. 4, and therefore the azimuthal angle should be fixed within a small range to extract the RABBIT phase or time-delay information. Otherwise, the time-delay information will be smeared out to the extent of averaging over $\phi$.
To decrease the systematic errors in CD spectra as much as possible, we regularly switched the chiral sample from $R$ to $S$ enantiomer for each group of comparative experiments under the same laser conditions. Importantly, the phase lock between XUV and IR beams was not interrupted when switching the sample, such that the measured chiral photoionisation time delay for two enantiomers was referenced to the same time zero. To accurately extract the forward-backward photoionisation time delay, we used Fourier transform to isolate the $2\omega$ components in the yield oscillation of SBs. Extended Data Fig. 6a shows the raw data of the SB 8 yield as a function of the XUV-IR delay, and the panel b displays the Fourier intensity spectrum of the panel a. An inverse Fourier transformation of the $2\omega$ oscillation signal is used to remove the DC background. The resulting data are shown in Fig. \ref{fig:figure3}d, which allows us to intuitively compare the phases of the yield oscillations. For the delay-resolved PECD shown in Fig. \ref{fig:figure2}d and Extended data Fig. \ref{fig:figure2}, we only remain the zero-frequency (DC) and $2\omega$ components and then preform the inverse Fourier transformation. We use the definition of 2[$I_S$($\tau$,$\theta$) - $I_R$($\tau$,$\theta$)]/[$I_S$($\tau$,$\theta$) + $I_R$($\tau$,$\theta$)] to calculate the time-resolved PECD, where $I_{S/R}$($\tau$,$\theta$) is the PAD after Fourier filtering at the XUV-IR delay of $\tau$ for the $S/R$ enantiomer. The energy integration window is 0.5 eV for each SB. For fitting the single-color $\theta$-resolved PECD, we use the first-order spherical harmonic function of $a_1\rm{cos}(\theta)$ to determine the asymmetry parameter $a_1$. For fitting the two-color $\theta$-resolved PECD, we used the fitting function $(a_1+a_2{\sin}^2(\theta)+a_3{\cos}^2(\theta)){\cos}(\theta)$ as derived in the Theoretical Methods section, which contains the higher-order terms $a_2$ an $a_3$. The PECD value at $\theta_k = 0$ is $a_1 + a_3$ in the two-color case and $a^{1\mathrm{ph}}_1$ in the one-color case. Therefore, we use $a_1 + a_3$ to represent the two-color PECD values shown in Figs. \ref{fig:figure2}c and \ref{fig:figure2}d of the main text and Extended Data Fig. 2. All angle-resolved and angle-integrated RABBIT traces shown in this work display the absolute electron counts. To generate the PECD plots, we normalized each PAD by its own maximum, because our laser and beamline stabilization might be not enough to resolve the absolute CD on the counts, which is a non-dipole effect and beyond our manuscript.

\subsection*{Theoretical methods}
\noindent\textbf{Computational model}.
For the simulation of the angle-resolved photoelectron spectra, we use the framework of Refs. \cite{goetz2019quantum,goetz2019perfect,goetz2024rabbitt}. Briefly, we describe the electronic states of the molecule in the frozen-core Hartree-Fock approximation \cite{lucchese1982photoionization,gianturco1994eps,natalense1999eps,greenman2017overset}, while treating the light-molecule interaction in the electric dipole approximation.
Hartree-Fock orbitals have been obtained with \texttt{MOLPRO} \cite{werner2012molpro,werner2012molprowires}.
The field-free continuum electronic states have been calculated with the \texttt{ePolyScat} program package \cite{gianturco1994eps,natalense1999eps,greenman2017overset}, using partial waves up to $L_{\mathrm{max}}=20$ for the single center expansion of the molecular potential which was found to yield sufficiently converged transition matrix elements for photoelectron energies up to 20 eV. In comparison, $L_{\mathrm{max}}=40$ has been used for CHBrClF in Ref.~\cite{goetz2024rabbitt}.
We use second-order time-dependent perturbation theory to simulate the photoionisation dynamics starting from the highest occupied molecular orbital. For the field intensities used in the experiment and the simulations, this is sufficient to describe the interference between the relevant ionisation pathways \cite{goetz2019quantum,goetz2019perfect,goetz2024rabbitt}.
A particular challenge is the calculation of transition matrix elements between electronic continuum states. Since application of the dipole operator localizes the perturbed ionised states, we can converge the representation of these states with respect to the Gaussian basis set, using the unoccupied orbitals from the Hartree-Fock calculation as intermediate states \cite{goetz2024rabbitt}. 
However, this approach has the caveat that the discretization of the continuum may introduce artificial resonances that strongly disturb the photoelectron spectrum. We use two strategies to circumvent this problem:
1. To achieve a dense sampling of the photoelectron continuum, we choose the large Gaussian basis set \texttt{aug-cc-pVQZ} \cite{kendall1992basis} and amend it by primitive diffuse $s$, $p$ and $d$ Gaussian functions with exponent coefficient of 0.015 atomic units centered at the center of charge of the molecule, as done in Ref.~\cite{goetz2017rempi}.
2. To remove residual artificial resonances, we fit the virtual Hartree-Fock energies and optimize the slope of the fitted energies, to minimize resonance effects.
The artificial resonances not only affect the intensity of the SB signals, but also strongly influence the PECD and the RABBITT traces. By choosing the slope of the fitted virtual energies such as to avoid resonance-like enhancements of individual SBs, we achieve good agreement with the experimentally found forward-backward delays shown in Fig.~\ref{fig:figure3}. Establishing a more detailed understanding of the effects of the artificial resonances is the subject of ongoing research.
The XUV pulse is modeled as a frequency comb comprising of the harmonics H7, H9, H11, H13 and H15 based on a fundamental frequency of 1.55 eV without a temporal delay. All XUV pulses have a Gaussian shaped intensity profile with FWHM of 5 fs.
An IR pulse oscillating at the fundamental frequency with the same intensity profile as the individual XUV pulses but two times their intensity has been applied without delay to the XUV pulses. The FWHM of the IR pulse intensity profile is 5 fs.
The parameters of the pulses are chosen to best reproduce the widths and relative intensities of the experimentally measured photoelectron spectrum shown in Fig.~\ref{fig:figure1}d.
Since the IR pulse used in the experiment is much longer than in the simulation, we fix the delay between the XUV and IR pulses and instead vary the carrier envelope phase of the IR pulse.

\noindent\textbf{Angle- and delay-depedendence of PECD}.
The perturbative treatment of the photoionisation results in the usual expressions for the orientation-averaged laboratory-frame PAD and PECD (see Refs.~\cite{goetz2019quantum,goetz2019perfect,goetz2024rabbitt} for details).
At the peaks in the photoelectron spectrum generated by the harmonics of the XUV comb, the PAD has the form
\begin{align*}
    \mathrm{PAD}_{\mathrm{H}}(k, \theta_k, \phi_k) =
    \sum_{L=0}^{2} \sum_{M=-L}^{L} \beta_{LM}^{1\mathrm{ph}}(k) \, Y_{LM}(\theta_k, \phi_k)
    \;.
\end{align*}
The one-photon expansion parameters, $\beta_{LM}^{1\mathrm{ph}}$, are independent of the XUV-IR delay. In case of a circularly polarized XUV, only the terms with $M=0$ are non-zero and the PECD takes, up to normalization, the well known form of
\begin{align*}
    \mathrm{PECD}_{\mathrm{H}}(k, \theta_k) = \beta_{10}^{1\mathrm{ph}}(k) \, \sqrt{\frac{3}{\pi}} \, \cos\theta_k
    = a^{1\mathrm{ph}}_1 \cos\theta_k
    \;,
\end{align*}
where $k$, $\theta_k$, and $\phi_k$ are the spherical polar coordinates of the photoelectron momentum.

At the SBs, the angular distribution for the pulse configurations used in the experiments reads
\begin{align}
    \mathrm{PAD}_{\mathrm{S}}(k, \theta_k, \phi_k) =
    \sum_{L=0}^{4} \sum_{M=0,\pm2} \beta_{LM}^{2\mathrm{ph}}(k) \, Y_{LM}(\theta_k, \phi_k)
    \;.
    \label{eq:pad_sb}
\end{align}
Under enantiomer exchange, the expansion parameters change sign according to $(-1)^L \beta_{LM}^{2\mathrm{ph}}$. Hence, the expression for the PECD at the SBs is up to normalization given by
\begin{align}
    \mathrm{PECD}_{\mathrm{S}}(k, \theta_k, \phi_k) =
    \sum_{L=1,3} \sum_{M=0,\pm2} 2 \beta_{LM}^{2\mathrm{ph}}(k) \, Y_{LM}(\theta_k, \phi_k)
    \;.
    \label{eq:pecd_sb}
\end{align}
When $\phi_k = 0$, this expression can be reduced to 
\begin{align*}
    \mathrm{PECD}_{\mathrm{S}}(k, \theta_k) =
\left(a_1 + a_2 \sin^2\theta_k + a_3 \cos^2\theta_k \right) \cos\theta_k
    \;,
\end{align*}
where the asymmetry parameters $a_1$, $a_2$ and $a_3$ are simple functions of $\beta_{LM}^{2\mathrm{ph}}$. The PECD value at $\theta_k = 0$ is $a_1 + a_3$ in the two-color case and $a^{1\mathrm{ph}}_1$ in the one-color case. Therefore, we use $a_1 + a_3$ to represent the two-color PECD values shown in Figs. \ref{fig:figure2}c and \ref{fig:figure2}d of the main text and Extended Data Fig. 1.

The two-photon anisotropy parameters at the SBs contain interfering contributions from the two ionisation pathways via the neighboring harmonics.
For co- or counter-rotating XUV and IR pulses, this pathway interference gives rise to the terms with $M=\pm2$, which oscillate as function of the XUV-IR delay with frequency $2\omega_{\mathrm{IR}}$, where $\omega_{\mathrm{IR}}$ is the frequency of the IR pulse. The $M=0$ terms only contain the single-pathway contributions and do not oscillate with the delay.
In contrast, an IR pulse linearly polarized in the polarization plane of the XUV pulse gives rise to both interfering and non-interfering contributions for both $M=0$ and $|M|=2$. These, however, differ in their delay dependence: for $M=0$, the interference terms oscillate with $2\omega_{\mathrm{IR}}$ whereas the single-pathway contributions do not oscillate. For $|M|=2$, the situation is reversed.
In the angle-resolved PECD, $M=0$ terms are maximal at $\theta_k=\SI{0}{\degree}$ and \SI{180}{\degree}, whereas terms with $|M|=2$ are maximal at $\theta_k\approx\SI{55}{\degree}$ and \SI{125}{\degree}.
Changing IR polarization and delay thus enables control over the anisotropy in the angle-resolved PECD.
Since both the PAD from Eq.~\eqref{eq:pad_sb} and the PECD from Eq.~\eqref{eq:pecd_sb} contain interference terms, the normalized PECD can show higher-order modulations of the $2\omega_{\mathrm{IR}}$ oscillations.
We extracted RABBIT traces and time delays from the angle-resolved photoelectron spectra with the same procedure as applied to the experimental data.

\printbibliography[notcategory=mainpart, title=\large{Methods references}]

@article{suda19a,
  title={Light-driven molecular switch for reconfigurable spin filters},
  author={Suda, Masayuki and Thathong, Yuranan and Promarak, Vinich and Kojima, Hirotaka and Nakamura, Masakazu and Shiraogawa, Takafumi and Ehara, Masahiro and Yamamoto, Hiroshi M},
  journal={Nature communications},
  volume={10},
  number={1},
  pages={2455},
  year={2019},
  publisher={Nature Publishing Group UK London}
}

@article{zhu21a,
  title={Multistate switching of spin selectivity in electron transport through light-driven molecular motors},
  author={Zhu, Qirong and Danowski, Wojciech and Mondal, Amit Kumar and Tassinari, Francesco and van Beek, Carlijn LF and Heideman, G Henrieke and Santra, Kakali and Cohen, Sidney R and Feringa, Ben L and Naaman, Ron},
  journal={Advanced science},
  volume={8},
  number={18},
  pages={2101773},
  year={2021},
  publisher={Wiley Online Library}
}

@article{ziv21a,
  title={Chirality nanosensor with direct electric readout by coupling of nanofloret localized plasmons with electronic transport},
  author={Ziv, Amir and Shoseyov, Omer and Karadan, Prajith and Bloom, Brian P and Goldring, Sharone and Metzger, Tzuriel and Yochelis, Shira and Waldeck, David H and Yerushalmi, Roie and Paltiel, Yossi},
  journal={Nano Letters},
  volume={21},
  number={15},
  pages={6496--6503},
  year={2021},
  publisher={ACS Publications}
}

@article{ritchie1976theory,
  title={Theory of the angular distribution of photoelectrons ejected from optically active molecules and molecular negative ions},
  author={Ritchie, Burke},
  journal={Physical Review A},
  volume={13},
  number={4},
  pages={1411},
  year={1976},
  publisher={APS}
}

@article{lucchese1982photoionization,
  title = {Studies of differential and total photoionization cross sections of molecular nitrogen},
  author = {Lucchese, Robert R. and Raseev, Georges and McKoy, Vincent},
  journal = {Physical Review A},
  volume = {25},
  pages = {2572--2587},
  year = {1982},
  publisher = {American Physical Society},
}

@article{kendall1992basis,
  title={Electron affinities of the first-row atoms revisited. Systematic basis sets and wave functions},
  author={Kendall, Rick A. and Dunning Jr., Thom H. and Harrison, Robert J.},
  journal={The Journal of Chemical Physics},
  volume={96},
  number={9},
  pages={6796-6806},
  year={1992},
  publisher={AIP}
}

@article{gianturco1994eps,
  author = {Gianturco, F. A. and Lucchese, R. R. and Sanna, N.},
  journal = {The Journal of Chemical Physics},
  number = {9},
  pages = {6464--6471},
  title = {Calculation of low‐energy elastic cross sections for electron‐CF$_4$ scattering},
  volume = {100},
  year = {1994}
}

@article{natalense1999eps,
  author = {Natalense, Alexandra P. P. and Lucchese, Robert R.},
  journal = {The Journal of Chemical Physics},
  number = {12},
  pages = {5344--5348},
  title = {Cross section and asymmetry parameter calculation for sulfur 1s photoionization of SF$_6$},
  volume = {111},
  year = {1999}
}

@article{Dorner2000,
  author = {D{\"{o}}rner, R. and Mergel, V. and Jagutzki, O. and Spielberger, L. and Ullrich, J. and Moshammer, R. and Schmidt-B{\"{o}}cking, H.},
  doi = {10.1016/S0370-1573(99)00109-X},
  journal = {Physics Reports},
  number = {2-3},
  pages = {95--192},
  title = {Cold Target Recoil Ion Momentum Spectroscopy: a `momentum microscope' to view atomic collision dynamics},
  volume = {330},
  year = {2000}
}

@article{bowering2001,
  title={Asymmetry in photoelectron emission from chiral molecules induced by circularly polarized light},
  author={B{\"o}wering, N and Lischke, Toralf and Schmidtke, B and M{\"u}ller, Norbert and Khalil, T and Heinzmann, Ulrich},
  journal={Physical Review Letters},
  volume={86},
  number={7},
  pages={1187},
  year={2001},
  publisher={APS}
}

@article{Ullrich2003,
  author = {Ullrich, J. and Moshammer, R. and Dorn, A. and D rner, R and Schmidt, L. Ph H. and Schmidt-B cking, H},
  doi = {10.1088/0034-4885/66/9/203},
  issn = {0034-4885},
  journal = {Reports on Progress in Physics},
  number = {9},
  pages = {1463--1545},
  title = {Recoil-ion and electron momentum spectroscopy: reaction-microscopes},
  volume = {66},
  year = {2003}
}

@article{zhao05a,
  title={Circularly-polarized laser-assisted photoionization spectra of argon for attosecond pulse measurements},
  author={Zhao, ZX and Chang, Zenghu and Tong, XM and Lin, CD},
  journal={Optics Express},
  volume={13},
  number={6},
  pages={1966--1977},
  year={2005},
  publisher={Optica Publishing Group}
}

@article{Lux2012,
  title={Circular dichroism in the photoelectron angular distributions of camphor and fenchone from multiphoton ionization with femtosecond laser pulses},
  author={Lux, Christian and Wollenhaupt, Matthias and Bolze, Tom and Liang, Qingqing and K{\"o}hler, Jens and Sarpe, Cristian and Baumert, Thomas},
  journal={Angewandte Chemie International Edition},
  volume={51},
  number={20},
  pages={5001--5005},
  year={2012},
  publisher={Wiley Online Library}
}

@Book{rodgers12a,
  title     = {Circular Dichroism: Theory and Spectroscopy},
  publisher = {Nova Science Publishers},
  year      = {2012},
  author    = {David S. Rodgers},
}

@misc{werner2012molpro,
  author = {H.-J. Werner and P. J. Knowles and G. Knizia and F. R. Manby and M. Sch{\"u}tz and P. Celani and T. Korona and R. Lindh and A. Mitrushenkov and G. Rauhut and      K. R. Shamasundar and T. B. Adler and R. D. Amos and A. Bernhardsson and A. Berning and D. L. Cooper and M. J. O. Deegan and A. J. Dobbyn and F. Eckert and E. Goll and C. Hampe     l and A. Hesselmann and G. Hetzer and T. Hrenar and G. Jansen and C. K{\"o}ppl and Y. Liu and A. W. Lloyd and R. A. Mata and A. J. May and S. J. McNicholas and W. Meyer and M.      E. Mura and A. Nicklass and D. P. O'Neill and P. Palmieri and D. Peng and K. Pfl{\"u}ger and R. Pitzer and M. Reiher and T. Shiozaki and H. Stoll and A. J. Stone and R. Tarroni      and T. Thorsteinsson and M. Wang},
  title = {MOLPRO, version 2012.1, a package of ab initio programs},
  url = {http://www.molpro.net}
}

@article{werner2012molprowires,
  author = {Werner, Hans-Joachim and Knowles, Peter J. and Knizia, Gerald and Manby, Frederick R. and Schütz, Martin},
  title = {Molpro: a general-purpose quantum chemistry program package},
  journal = {Wiley Interdisciplinary Reviews: Computational Molecular Science},
  volume = {2},
  number = {2},
  pages = {242-253},
  year = {2012},
}

@article{Lehmann2013,
  title={Imaging photoelectron circular dichroism of chiral molecules by femtosecond multiphoton coincidence detection},
  author={Lehmann, C Stefan and Ram, N Bhargava and Powis, Ivan and Janssen, Maurice HM},
  journal={The Journal of Chemical Physics},
  volume={139},
  number={23},
  year={2013},
  publisher={AIP Publishing}
}

@article{garcia2014photoelectron,
  title={Photoelectron circular dichroism and spectroscopy of trifluoromethyl-and methyl-oxirane: a comparative study},
  author={Garcia, Gustavo A and Dossmann, Heloise and Nahon, Laurent and Daly, Steven and Powis, Ivan},
  journal={Physical Chemistry Chemical Physics},
  volume={16},
  number={30},
  pages={16214--16224},
  year={2014},
  publisher={Royal Society of Chemistry}
}

@article{zipp2014,
  title={Probing electron delays in above-threshold ionization},
  author={Zipp, Lucas J and Natan, Adi and Bucksbaum, Philip H},
  journal={Optica},
  volume={1},
  number={6},
  pages={361--364},
  year={2014},
  publisher={Optical Society of America}
}

@article{Cireasa2015,
  title={Probing molecular chirality on a sub-femtosecond timescale},
  author={Cireasa, Raluca and Boguslavskiy, AE and Pons, B and Wong, MCH and Descamps, D and Petit, S and Ruf, H and Thir{\'e}, Nicolas and Ferr{\'e}, A and Suarez, J and Higuet, J.and Schmidt, B. E. and Alharbi, A. F. and L{\'e}gar{\'e}, F. and Blanchet, V. and Fabre, B. and Patchkovskii, S. and Smirnova, O. and Mairesse, Y. and Bhardwaj, V. R.},
  journal={Nature Physics},
  volume={11},
  number={8},
  pages={654--658},
  year={2015},
  publisher={Nature Publishing Group UK London}
}

@article{Lux2015,
  title={Photoelectron circular dichroism of bicyclic ketones from multiphoton ionization with femtosecond laser pulses},
  author={Lux, Christian and Wollenhaupt, Matthias and Sarpe, Cristian and Baumert, Thomas},
  journal={ChemPhysChem},
  volume={16},
  number={1},
  pages={115--137},
  year={2015},
  publisher={Wiley Online Library}
}

@article{Beaulieu2017,
  author = {S. Beaulieu and A. Comby and A. Clergerie and J. Caillat and D. Descamps and N. Dudovich and B. Fabre and R. Géneaux and F. Légaré and S. Petit and B. Pons and G. Porat and T. Ruchon and R. Taïeb and V. Blanchet and Y. Mairesse},
  doi = {10.1126/science.aao5624},
  journal = {Science},
  pages = {1288-1294},
  pmid = {29217568},
  title = {Attosecond-resolved photoionization of chiral molecules},
  volume = {358},
  year = {2017},
}

@article{goetz2017rempi,
  author = {Goetz, R. E. and Isaev, T. A. and Nikoobakht, B. and Berger, R. and Koch, C. P.},
  title = {Theoretical description of circular dichroism in photoelectron angular distributions of randomly oriented chiral molecules after multi-photon photoionization},
  journal = {The Journal of Chemical Physics},
  volume = {146},
  number = {2},
  pages = {024306},
  year = {2017},
}

@article{greenman2017overset,
  author = {Greenman, Loren and Lucchese, Robert R. and McCurdy, C. William},
  journal = {Physical Review A},
  number = {5},
  pages = {052706},
  title = {Variational treatment of electron–polyatomic-molecule scattering calculations using adaptive overset grids},
  volume = {96},
  year = {2017}
}

@Article{Baykusheva2018,
  author    = {Baykusheva, Denitsa and W\"orner, Hans Jakob},
  title     = {Chiral Discrimination through Bielliptical High-Harmonic Spectroscopy},
  journal   = {Physical Review X},
  year      = {2018},
  volume    = {8},
  pages     = {031060},
  numpages  = {12},
  publisher = {American Physical Society},
}

@Article{beaulieu18a,
  author    = {Beaulieu, Samuel and Comby, Antoine and Descamps, Dominique and Fabre, Baptiste and Garcia, GA and G{\'e}neaux, Romain and Harvey, AG and L{\'e}gar{\'e}, Francois and Ma{\v{s}}{\'i}n, Z and Nahon, Laurent and others},
  title     = {Photoexcitation circular dichroism in chiral molecules},
  journal   = {Nature Physics},
  year      = {2018},
  volume    = {14},
  number    = {5},
  pages     = {484--489},
  publisher = {Nature Publishing Group},
}

@article{demekhin18a,
  title={Photoelectron circular dichroism with two overlapping laser pulses of carrier frequencies $\omega$ and 2 $\omega$ linearly polarized in two mutually orthogonal directions},
  author={Demekhin, Philipp V and Artemyev, Anton N and Kastner, Alexander and Baumert, Thomas},
  journal={Physical Review Letters},
  volume={121},
  number={25},
  pages={253201},
  year={2018},
  publisher={APS}
}

@article{Baykusheva2019,
  title={Real-time probing of chirality during a chemical reaction},
  author={Baykusheva, Denitsa and Zindel, Daniel and Svoboda, V{\'\i}t and Bommeli, Elias and Ochsner, Manuel and Tehlar, Andres and W{\"o}rner, Hans Jakob},
  journal={Proceedings of the National Academy of Sciences of the United States of America},
  volume={116},
  number={48},
  pages={23923--23929},
  year={2019},
  publisher={National Acad Sciences}
}

@article{goetz2019perfect,
  title={Perfect control of photoelectron anisotropy for randomly oriented ensembles of molecules by XUV REMPI and polarization shaping},
  author={Goetz, R Esteban and Koch, Christiane P and Greenman, Loren},
  journal={The Journal of Chemical Physics},
  volume={151},
  number={7},
  year={2019},
  publisher={AIP Publishing}
}

@article{goetz2019quantum,
  title={Quantum control of photoelectron circular dichroism},
  author={Goetz, R Esteban and Koch, Christiane P and Greenman, Loren},
  journal={Physical Review Letters},
  volume={122},
  number={1},
  pages={013204},
  year={2019},
  publisher={APS}
}

@Article{naaman19a,
  author    = {Naaman, Ron and Paltiel, Yossi and Waldeck, David H},
  title     = {Chiral molecules and the electron spin},
  journal   = {Nature Reviews Chemistry},
  year      = {2019},
  volume    = {3},
  number    = {4},
  pages     = {250--260},
  publisher = {Nature Publishing Group},
}

@article{rozen19a,
  title={Controlling subcycle optical chirality in the photoionization of chiral molecules},
  author={Rozen, Shaked and Comby, Antoine and Bloch, Etienne and Beauvarlet, Sandra and Descamps, Dominique and Fabre, Baptiste and Petit, Stephane and Blanchet, Valerie and Pons, Bernard and Dudovich, Nirit and Mairesse, Yann},
  journal={Physical Review X},
  volume={9},
  number={3},
  pages={031004},
  year={2019},
  publisher={APS}
}

@article{Ji2021,
   author = {Jia-Bao Ji and Saijoscha Heck and Meng Han and Hans Jakob Wörner},
   doi = {10.1364/OE.432222},
   journal = {Optics Express},
   pages = {27732},
   title = {Quantitative uncertainty determination of phase retrieval in RABBITT},
   volume = {29},
   year = {2021},
}

@Article{wu21a,
  author    = {Wu, Yanze and Subotnik, Joseph E},
  title     = {Electronic spin separation induced by nuclear motion near conical intersections},
  journal   = {Nature Communications},
  year      = {2021},
  volume    = {12},
  number    = {1},
  pages     = {700},
  publisher = {Nature Publishing Group UK London},
}

@Article{evers22a,
  author = {Evers, Ferdinand and Aharony, Amnon and Bar-Gill, Nir and Entin-Wohlman, Ora and Hedegård, Per and Hod, Oded and Jelinek, Pavel and Kamieniarz, Grzegorz and Lemeshko, Mikhail and Michaeli, Karen and Mujica, Vladimiro and Naaman, Ron and Paltiel, Yossi and Refaely-Abramson, Sivan and Tal, Oren and Thijssen, Jos and Thoss, Michael and van Ruitenbeek, Jan M. and Venkataraman, Latha and Waldeck, David H. and Yan, Binghai and Kronik, Leeor},
  title     = {Theory of chirality induced spin selectivity: Progress and challenges},
  journal   = {Advanced Materials},
  year      = {2022},
  volume    = {34},
  number    = {13},
  pages     = {2106629},
  publisher = {Wiley Online Library},
}

@Article{matselyukh22a,
  author    = {Matselyukh, Danylo T and Despr{\'e}, Victor and Golubev, Nikolay V and Kuleff, Alexander I and W{\"o}rner, Hans Jakob},
  title     = {Decoherence and revival in attosecond charge migration driven by non-adiabatic dynamics},
  journal   = {Nature Physics},
  year      = {2022},
  volume    = {18},
  number    = {10},
  pages     = {1206--1213},
  publisher = {Nature Publishing Group},
}

@Article{bloom24a,
  title={Chiral induced spin selectivity},
  author={Bloom, Brian P and Paltiel, Yossi and Naaman, Ron and Waldeck, David H},
  journal={Chemical Reviews},
  volume={124},
  number={4},
  pages={1950--1991},
  year={2024},
  publisher={ACS Publications}
}

@article{Svoboda2022,
  title={Femtosecond photoelectron circular dichroism of chemical reactions},
  author={Svoboda, V{\'\i}t and Ram, Niraghatam Bhargava and Baykusheva, Denitsa and Zindel, Daniel and Waters, Max DJ and Spenger, Benjamin and Ochsner, Manuel and Herburger, Holger and Stohner, J{\"u}rgen and W{\"o}rner, Hans Jakob},
  journal={Science Advances},
  volume={8},
  number={28},
  pages={eabq2811},
  year={2022},
  publisher={American Association for the Advancement of Science}
}

@article{Han2023metrology,
  title={Attosecond metrology in circular polarization},
  author={Han, Meng and Ji, Jia-Bao and Ueda, Kiyoshi and W{\"o}rner, Hans Jakob},
  journal={Optica},
  volume={10},
  number={8},
  pages={1044--1052},
  year={2023},
  publisher={Optica Publishing Group}
}

@article{Han2023attosecond,
  title={Attosecond circular-dichroism chronoscopy of electron vortices},
  author={Han, Meng and Ji, Jia-Bao and Bal{\v{c}}i{\=u}nas, Tadas and Ueda, Kiyoshi and W{\"o}rner, Hans Jakob},
  journal={Nature Physics},
  volume={19},
  number={2},
  pages={230--236},
  year={2023},
  publisher={Nature Publishing Group UK London}
}

@article{han2024separation,
  title={Separation of photoionization and measurement-induced delays},
  author={Han, Meng and Ji, Jia-Bao and Leung, Chung Sum and Ueda, Kiyoshi and W{\"o}rner, Hans Jakob},
  journal={Science Advances},
  volume={10},
  number={4},
  pages={eadj2629},
  year={2024},
  publisher={American Association for the Advancement of Science}
}

@article{goetz2024rabbitt,
  title={Continuum-electron interferometry for enhancement of photoelectroncircular dichroism and measurement of bound, free, and mixed contributions to chiral response},
  author={Goetz, R Esteban and Blech, Alexander and Allison, Corbin and Koch, Christiane P and Greenman, Loren},
  journal={arXiv:2104.07522}
}

\clearpage

\begin{extendedFig}[t]
\centering
\includegraphics[width=\linewidth]{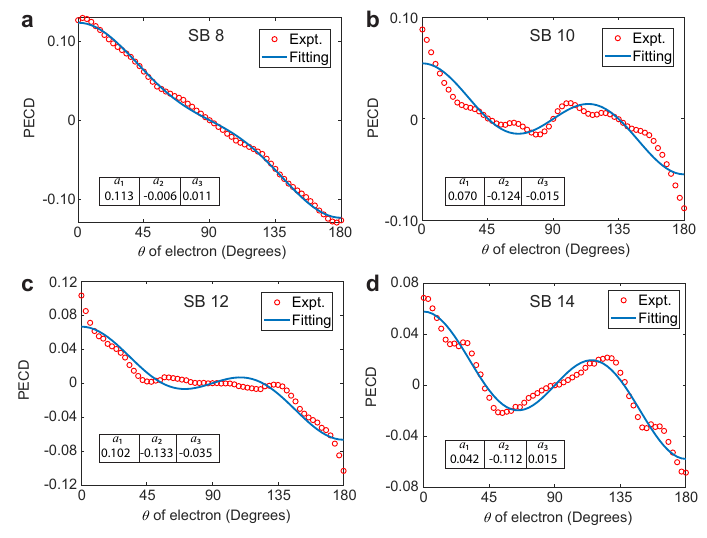}
\caption{{\bf Determination of asymmetry parameters in two-color fields.} \textbf{a-d}, Angle-resolved PECD distributions at SB positions, which are extracted from Fig. \ref{fig:figure2}b. By fitting the angle-resolved PECD distributions with the two-color formula shown in the Theoretical Methods section, we can extract the two-color two-photon asymmetry parameters in the laboratory frame, which are shown in the right lower corner of the panels.}
\label{fig:ED1}
\end{extendedFig}

\begin{extendedFig}[t]
\centering
\includegraphics[width=7 cm]{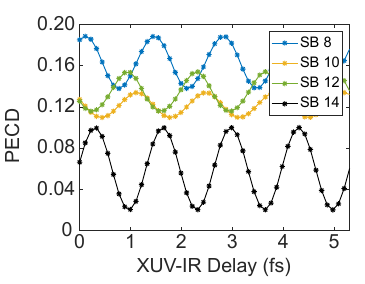}
\caption{{\bf Delay-resolved PECD of SBs in the circularly polarized, co-rotating IR and XUV fields.}}
\label{fig:ED2}
\end{extendedFig}

\begin{extendedFig}[t]
\centering
\includegraphics[width=\linewidth]{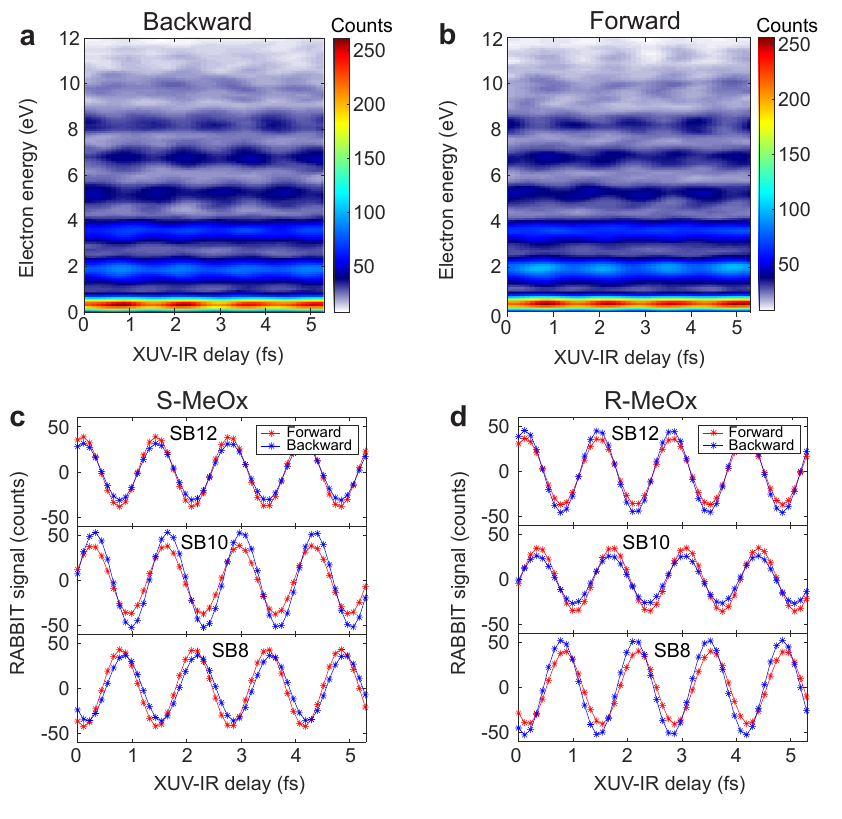}
\caption{\textbf{Experimental data for S enantiomer in the linearly polarized IR field}. \textbf{a, b}, Angle-integrated RABBIT traces for backward and forward photoelectrons from the S enantiomer, respectively. \textbf{c}, The extracted yield oscillations of the SBs from a and b. \textbf{d}, The corresponding results from the R enantiomer, which is already shown in Fig. \ref{fig:figure3}d }
\label{fig:ED3}
\end{extendedFig}

\begin{extendedFig}[t]
\centering
\includegraphics[width=\linewidth]{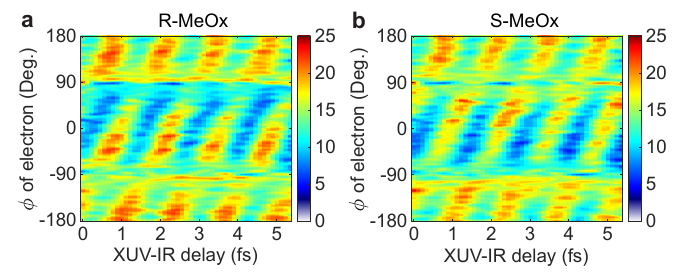}
\caption{\textbf{$\phi$-resolved RABBIT trace}. \textbf{a,b}, $\phi$-resolved RABBIT traces of SB 8 in the circularly polarized IR field for the $R$- and $S$-enantiomers of MeOx, respectively. Due to the angular streaking effect, photoelectrons at different $\phi$ angles will have different RABBIT phases. Therefore, the $\phi$ angle of photoelectron must be resolved to correctly measure the photoionisation time delay. Note that there is no angular resolution at $\phi = \pm 90^\circ$ due to the static magnetic field used in the COLTRIMS spectrometer.}
\label{fig:ED4}
\end{extendedFig}

\begin{extendedFig}[t]
\centering
\includegraphics[width=\linewidth]{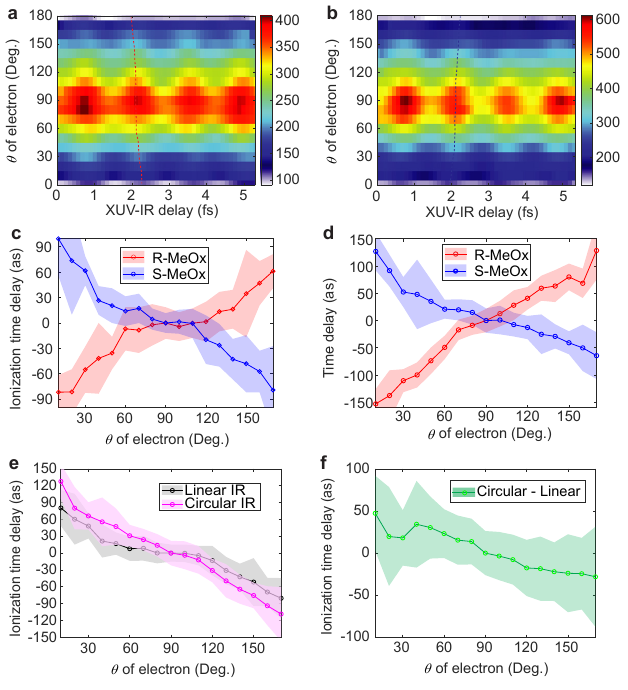}
\caption{\textbf{Two-color angle-resolved photoionisation time delay}. \textbf{a,b}, $\theta$-resolved RABBIT traces of SB 8 in the linearly polarized IR field for the $R$- and $S$-enantiomers of MeOx, respectively. \textbf{c}, The extracted photoionisation time delay from a and b, where the delay values at $\theta = 90^\circ$ were chosen as the zero point. \textbf{d}, The corresponding photoionisation time delay in the circularly polarized IR field, which is already shown in Fig. \ref{fig:figure4}c. \textbf{e}, The symmetrized time delay between two enantiomers, i.e., $(\tau(\theta)_{S-\rm{MeOx}} + \tau(180^{\circ}-\theta)_{R-\rm{MeOx}})/2$. \textbf{d}, Time delay difference between circular and linear polarizations, isolating the chiral part of the cc delays.}
\label{fig:ED5}
\end{extendedFig}

\begin{extendedFig}[t]
\centering
\includegraphics[width=\linewidth]{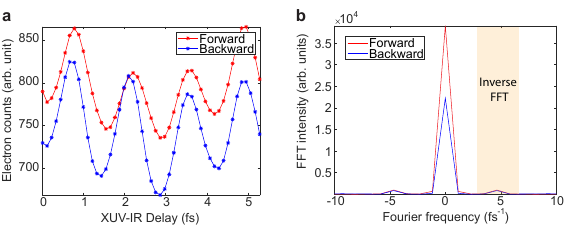}
\caption{\textbf{Fourier analysis of the experimental data}. \textbf{a}, Raw data of the SB 8 yield as a function of the XUV-IR delay. \textbf{b}, Fourier intensity spectrum of the data shown in a, where the range of the 2$\omega$ oscillation signal has been highlighted in orange. An inverse Fourier transformation of the 2$\omega$ oscillation signal is used to remove the DC background. The resulting data are shown in Fig. \ref{fig:figure3}d.}
\label{fig:ED6}
\end{extendedFig}

\end{document}